\documentclass[preprint,nofootinbib,showpacs,aps,prc]{revtex4-1}
\usepackage{graphicx}
\usepackage{amstext}
\usepackage{amssymb}

\usepackage[usenames]{color}
\begin{document}
\title{Femtoscopy correlations of kaons in $Pb + Pb$ collisions at LHC within hydrokinetic model}
\author{V.M. Shapoval$^a$, P. Braun-Munzinger$^b$, Iu.A. Karpenko$^{a,c}$  }
\author{Yu.M. Sinyukov$^{a,b}$}
\affiliation{$(a)$ Bogolyubov Institute for Theoretical Physics, Metrolohichna str. 14b, 03680 Kiev, Ukraine \\ $(b)$ ExtreMe Matter Institute EMMI, GSI~Helmholtz~Zentrum f\"ur~Schwerionenforschung,
D-64291 Darmstadt, Germany\\ $(c)$ Frankfurt~Institute~for~Advanced~Studies, Ruth-Moufang-Str.~1, 60438 Frankfurt am Main, Germany}

\begin{abstract}
We provide, within the hydrokinetic model, a detailed investigation of kaon interferometry in $Pb+Pb$ collisions at LHC energy ($\sqrt{s_{NN}} = 2.76$ TeV). Predictions are presented for 1D interferometry radii of $K^0_SK^0_S$ and $K^{\pm}K^{\pm}$ pairs as well as for 3D femtoscopy scales in {\it out}, {\it side} and {\it long} directions. The results are compared with existing pion interferometry radii. We also make predictions for full LHC energy. 
\end{abstract}

 \maketitle



\section{Introduction}
Correlation femtoscopy \cite{Kopylov} is a tool to study the spatiotemporal structure of particle emission in nucleus-nucleus, proton-(anti)proton and proton-nucleus collisions. This structure is correlated with the dynamics of the collision processes \cite{Pratt, MakSin, Sinyuk, Hama} which can hence be studied with interferometry tools. The corresponding femtoscopic patterns can be presented in different forms. One is the $k_T$-momentum dependence of the interferometry radii $R_i(k_T=(\left|{\bf p}_{T1}+{\bf p}_{T2}\right|)/2)$, that results from a 3D Gaussian fit in $q_i=p_{1i}-p_{2i}$ of the two-particle  correlation function $C({\bf q}, k_T)$, defined as a ratio of the two-particle spectrum to the product of the single-particle ones. The other one is the source function $S({\bf r^*})$ reflecting the dependence of the pair production on the distance ${\bf r^*}$ between the two emitted particles in the rest frame of the pair. Both patterns supplement each other, and a reliable model should describe/predict all the mentioned types of the  femtoscopic observables, if it contains a detailed space-time picture of the collision process.

It is important to note that the correlation function behavior depends also on the particle species. The detailed behavior of this dependence can can be used to discriminate between different scenarios of the matter evolution and particle emission in the collision processes. For example, the hydrodynamic picture of A+A collisions for the particular case of negligible transverse flow leads to the same $m_T^{-1/2}$ behavior of the longitudinal radii $R_i(k_T)$ for identical pions and kaons, and even gives the complete  $m_T$-scaling in the case of common freeze-out \cite{MakSin,Sinyuk}\footnote{Here, $m_T^2=m^2+((p_{1T}+p_{2T})/2)^2$ is the transverse mass of the particle pair.}. In simple analytical models a deviation from such a scaling can be a signal of enhanced transverse flow \cite{AkkSin} and/or different (effective) freeze-out times, e.g., because kaons are less affected by the decay of resonances than pions at the afterburner stage. The last factor could, in principle, also affect  differences in femtoscopic scales between charged identical kaon pairs and $K^0_SK^0_S$, and  corresponding theoretical estimates and comparison with experimental data are to the point here. Note that, in spite of analytical approximations, in realistic hybrid or hydrokinetic models many factors act simultaneously and the results can be obtained only by time-consuming  numerical calculations.

The hydrokinetic model (HKM) \cite{HKM,KS} was developed to describe simultaneously a wide class of bulk observables in A+A collisions at top 
RHIC and LHC energies, to predict pion, kaon, proton, as well as all charged, particle spectra for all centralities, $v_2$ coefficients and pion femtoscopy scales \cite{hHKM}. Also the pion and kaon source functions at the same initial conditions were well described at top RHIC energy, and predictions for LHC were done \cite{ShapSin}.
In addition, HKM well describes pion interferometry radii in p+p collision at LHC ($\sqrt{s} = 7$ TeV) energy
if one incorporates the quantum uncertainty principle into a quasi-classical event generator \cite{PBM-Sin}.

In this work we apply HKM to workout and predict different kaon femtoscopy scales at LHC energy in Pb+Pb ($\sqrt{s_{NN}} = 2.76$ TeV) collisions at the same parameter values as those used in \cite{hHKM}. The predictions for kaon and pion interferometry radii at full LHC energy ($\sqrt{s_{NN}} = 5.12$ TeV) are demonstrated also.
The kaon and pion femtoscopy analysis gives the possibility to clarify whether the
resulting spatiotemporal structure of emission functions for different particle species, that describe well the bulk observables at LHC, is self-consistent and reliable. Then this picture of particle emission will serve as a reference point no matter how much advanced future models will be developed.

\section{Hydrokinetic description of A+A collisions} 

The hydrokinetic model \cite{HKM,KS} was developed to simulate the evolution of matter formed in  relativistic heavy-ion collisions.
The full process proceeds through stages -- a high density medium expansion, described in the ideal hydrodynamics approximation, then gradual system decoupling, described in the hydrokinetic
approach. The final stage is a hadronic cascade within UrQMD. At the first stage matter is assumed to be in local chemical and thermal equilibrium. Here we use a lattice-QCD inspired equation of state for the quark-gluon phase \cite{qcd},  matched via a cross-over type transition with the hadron resonance gas, consisting of all 329 well-established hadron states made out of u, d, s quarks.
As the system expands and cools down, it reaches the second stage, which begins at the chemical freeze-out isotherm $T_{ch}=165$~MeV \cite{Andronic}\footnote{Note that the last analysis gives the chemical freeze-out temperature $T=156$ MeV \cite{Stachel} for LHC energy $\sqrt{s_{NN}}=2.76$ TeV in thermal model that ignores inelastic processes at afterburner stage except decays of resonances.}.  At temperatures $T<T_{ch}$ system gradually falls out of both chemical and thermal equilibrium,
and the particles begin to continuously escape from the medium. 
In the hybrid model version (hHKM) \cite{hHKM} the hydrokinetic description of the second stage is switched to the UrQMD hadron cascade on a space-like hypersurface, situated behind the  hadronization phase. Another option is a direct switching to the cascade just from the hydrodynamic stage, at the hadronization hypersurface $T_{ch}=165$~MeV. 
We use this particular variant in the current analysis, relying on the result of \cite{hHKM}, where the comparison with data of one- and two-particle spectra, calculated at both types of matching hydro and cascade stages, showed a rather small
difference between them in the considered case of top RHIC and LHC energies. 
The reason for the similarity is that, for the utilized event-averaged initial conditions, the contribution from the loss of particles crossing non-space-like sectors which match the hydro-UrQMD hypersurface is quite small, $\sim 1-2$~percent. This is related to the  very high velocities ($0.7c$) of the fluid elements crossing non-space-like parts of the chemical freeze-out isotherm. Then the number of the particles that move inside the fluid belongs to a tail of the relativistic (Boltzmann) spectra and their negative contributions in the Cooper-Frye formula \cite{Cooper} are negligible. 

At the switching hypersurface a set of particles is generated according to the chemical freeze-out distribution function \cite{Andronic}
using either Cooper-Frye prescription \cite{Cooper} (for sudden switching from hydro to UrQMD) or using the technique of Boltzmann equations in integral form \cite{HKM} (if hydrokinetics is involved). This set serves as input for UrQMD \cite{urqmd}, within which  particles rescatter and decay. The final model output is again a collection of particles, characterized by their momenta and the points of their last collisions.

\begin{figure}[!hbt]
\includegraphics[width=0.9\textwidth]{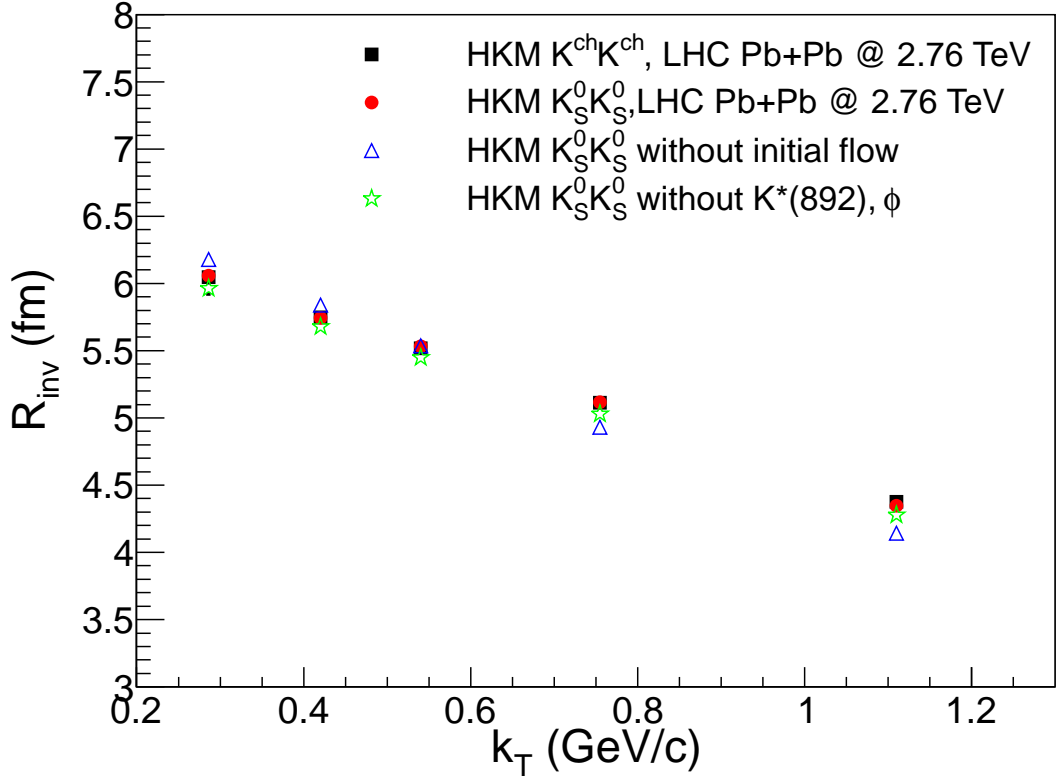}
\caption{The HKM prediction for the dependence of $K^{ch}K^{ch}$ and 
$K_S^{0}K_S^{0}$ interferometry radii $R_{inv}$ on $k_T$ for $\sqrt{s_{NN}}=2.76$~TeV Pb+Pb LHC collisions, $c=0-5\%$, $|\eta|<0.8$, $0.14<p_T<1.5$~GeV/$c$.} 
\label{r1d}
\end{figure}

We work in the central rapidity slice and assume longitudinal boost-invariance. This is well fulfilled at LHC energy \cite{Alice}. Early thermalization at proper time $\tau = 0.1$ fm/c is assumed. In the transverse plane we use Glauber Monte Carlo initial energy density profile generated in the GLISSANDO code \cite{glissando}. Fluctuations of the initial conditions
tilt in each event the principal axes of the ellipse of inertia and
shift the center of mass relative to the reaction-plane coordinate
system. To account for this effect, we superimpose the principal
axes by rotation and recentering of each initial distribution and
after that take averages over the ensemble of events (so-called
variable geometry analysis, also implemented as an option in the GLISSANDO code). So we use event-averaged initial
conditions. 
We assume zero and small but non-zero initial transverse flow which is taken linear in transverse radius $r_T$ \cite{hHKM}: $y_T = \alpha \frac{r_T}{R^2(\phi)}$. Here $R(\phi)$ is the system's homogeneity length in 
$\phi$-direction, we take it as the r.m.s. $R(\phi)=\sqrt{\left\langle  r^2 \right\rangle_{\phi}}$ along the azimuthal angle $\phi$. Such a small initial flow mimics  shear viscosity effects during the system hydrodynamic evolution 
as well as effects of event by event fluctuating hydro-solutions \cite{hHKM}.  
The maximal initial energy density $\epsilon_0$ is chosen to reproduce the experimental (or predicted for full LHC energy) mean charged particle multiplicity. Thus, $\epsilon_0$ 
and the coefficient $\alpha$ are the only fitting parameters of the model which are attributed to the initial time 0.1  fm/c. We take
the parameters from \cite{hHKM} that provide the best fit for the charged particle multiplicity, pion, kaon and proton spectra and pion interferometry data, 
$\epsilon_0 = 1300$~GeV/fm$^3$ and $\alpha = 0.45$~fm (the maximal initial transverse velocity at the very periphery of the system is then 0.05). We also demonstrate the results with no initial transverse flow.

\begin{figure}[!hbt]
\includegraphics[width=0.9\textwidth]{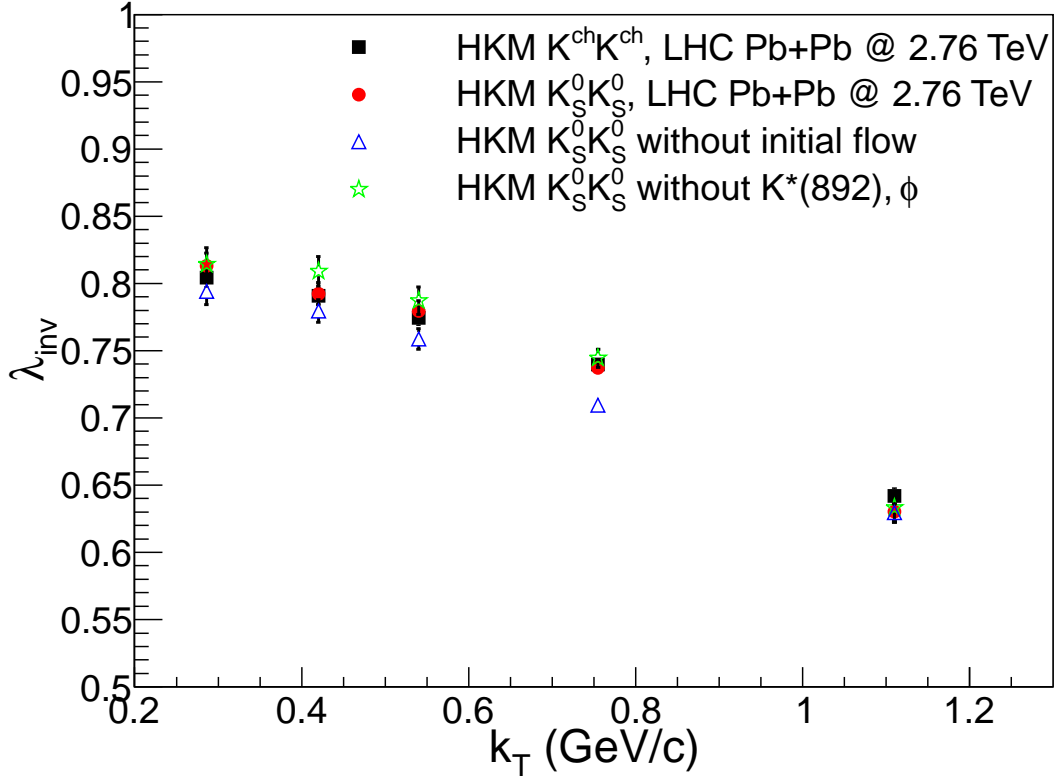}
\caption{HKM prediction for the dependence of $K^{ch}K^{ch}$ and 
$K_S^{0}K_S^{0}$  parameter $\lambda_{inv}$ on $k_T$ for $\sqrt{s}=2.76$~GeV Pb+Pb LHC collisions, $c=0-5\%$, $|\eta|<0.8$, $0.14<p_T<1.5$~GeV/$c$.} 
\label{lam1d}

\end{figure}

\section{Results and discussion}
At the initial conditions described in the previous section we calculate within HKM the interferometry radii $R_{inv}$ for  charged and neutral kaon correlation functions $C(q_{inv},k_T)$ where $k_T$ is the absolute value of half-momentum of the pair. The kaon pairs are generated in central $(c=0-5 \%)$ LHC Pb+Pb collisions at the energy $\sqrt{s_{NN}}=2.76$~TeV. The particles with transverse momentum in the range $0.14<p_T<1.5$~GeV/$c$ and pseudorapidity $|\eta|<0.8$ were chosen for the analysis. The results of the Gaussian fits, $C(q_{inv}, k_T)=1+\lambda_{inv}(k_T) \exp(-q^2_{inv}/R^2_{inv}(k_T)$, are presented in Fig. \ref{r1d}. One can see there also the interferometry radii for the case without initial transverse flow, $\alpha = 0$. Since $C(q_{inv}, k_T)$ is a non-Gaussian function, the parameter $\lambda_{inv}$ is small comparing with the intercept of the correlation function and it decreases with $k_T$ as one can see in Fig. \ref{lam1d}. We demonstrate in addition the $k_T$-behavior of $\lambda_{inv}$ when there is no initial transverse flow and also for the artificial case when the resonances $K^*(892)$ and $\phi (1020)$ decay just on the hypersurface of the chemical freeze-out. As one can see such an "exclusion"' of the resonances almost does not affect the femtoscopy scales. That is the main reason why the interferometry radii for charged and neutral kaons practically coincide.   
\begin{figure}[!hbt]
\center
\includegraphics[width=1.08\textwidth]{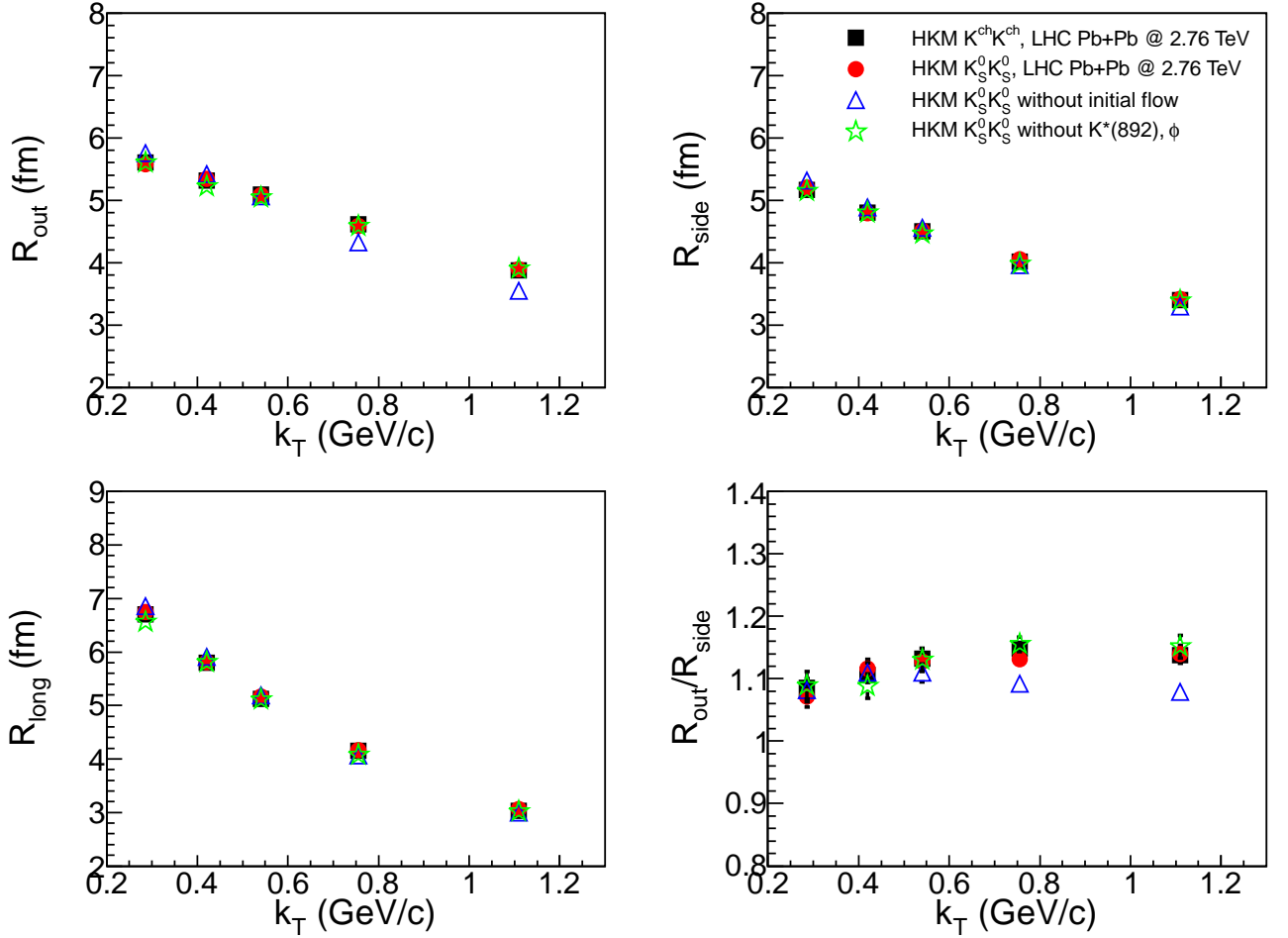}
\caption{HKM predictions for the dependence of $K^{ch}K^{ch}$ and 
$K_S^{0}K_S^{0}$ 3D interferometry radii $R_{i}$ on $k_T$ for $\sqrt{s}=2.76$~GeV Pb+Pb LHC collisions, $c=0-5\%$, $|\eta|<0.8$, $0.14<p_T<1.5$~GeV/$c$.} 
\label{kk3d}
\end{figure}

\begin{figure}[!hbt]
\center
\includegraphics[width=0.9\textwidth]{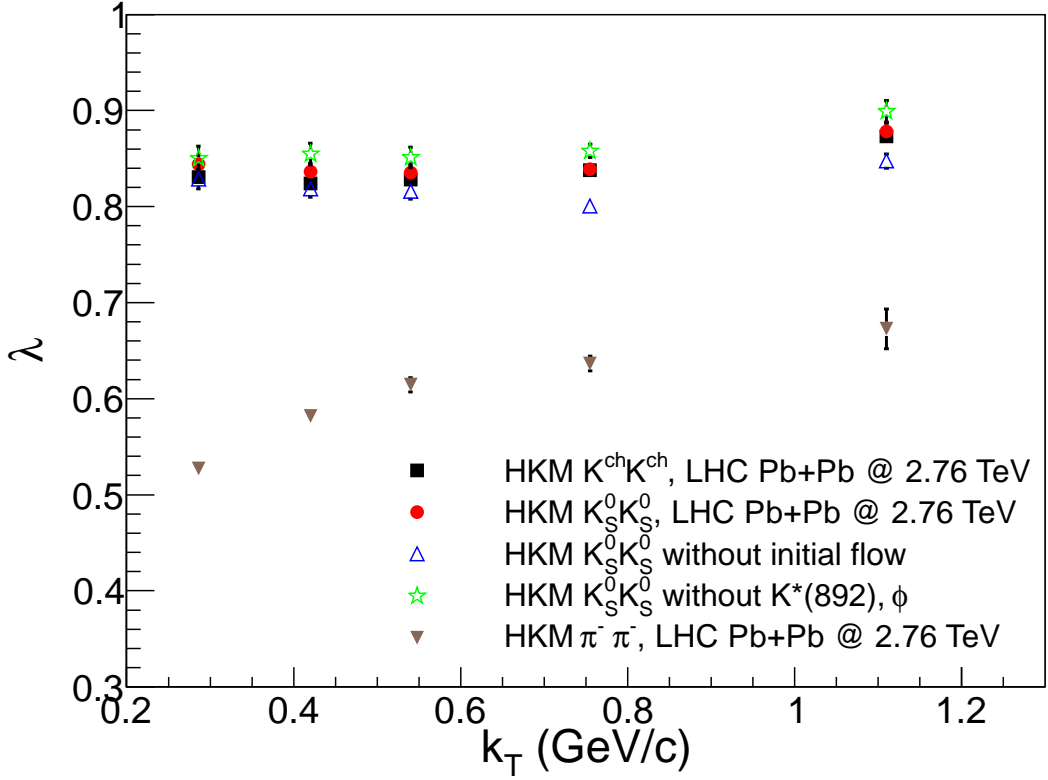}
\caption{Comparison of the HKM $k_T$-dependencies for $K^{ch}K^{ch}$, $K_S^{0}K_S^{0}$ and $\pi^{-} \pi^{-}$ parameter $\lambda$ for $\sqrt{s_{NN}}=2.76$~TeV Pb+Pb LHC collisions, $c=0-5\%$, $|\eta|<0.8$, $0.14<p_T<1.5$~GeV/$c$.} 
\label{kklam}
\end{figure}

\begin{figure}[phbt]
\center
\includegraphics[width=1.08 \textwidth]{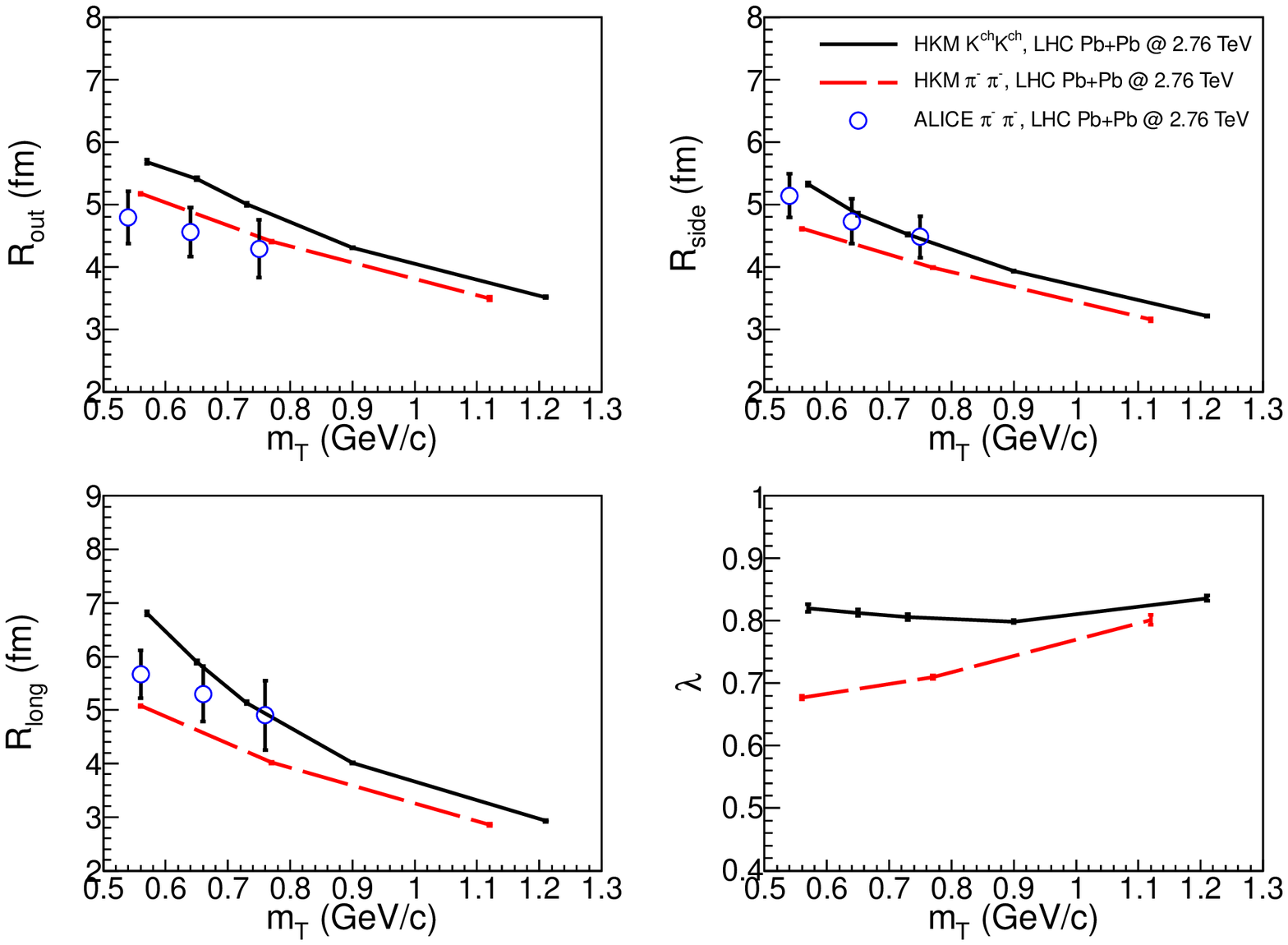}
\caption{Comparison of the $m_T$-dependence of $K^{ch}K^{ch}$ and $\pi^{-}\pi^{-}$ 3D interferometry radii for $\sqrt{s_{NN}}=2.76$~TeV Pb+Pb LHC collisions, $c=0-5\%$, $|\eta|<0.8$, $0.14<p_T<1.5$~GeV/$c$.} 
\label{k-pi}
\end{figure}

In Fig. \ref{kk3d} we show plots of the $k_T$-dependence 
of 3D $R_{i}$ interferometry radii and of the ratio $R_{out}$ to $R_{side}$. These radius parameters are extracted from Gaussian fits to $K^{ch}K^{ch}$ and $K_S^{0}K_S^{0}$ correlation function histograms calculated in the hydrokinetic model. The dependence on $k_T$ of the kaon suppression parameter $\lambda$ is demonstrated in Fig. \ref{kklam} in comparison with the corresponding result for pions. Opposite to the case of kaons, the contribution of pions from long-lived resonance decays is significant, so the suppression parameter $\lambda$ is significantly less for pions and it noticeably grows with $k_T$ since the contribution from these resonances is reduced with $k_T$ because of kinematics.  

In addition, in Fig. \ref{k-pi} we compare the $m_T$ behavior of $R_{i}$ for charged identical kaons with that the model gives for identical charged pions. We see that the hydrokinetic model does not result in  $m_T$-scaling  for the pion and kaon {\it side-} interferometry radii, with kaon radii  larger than values for pions at the same $m_T$. The deviation from the scaling behavior is especially significant in the {\it long-} direction that is a consequence of strong transverse flow \cite{AkkSin}. 

\begin{figure}[phbt]
\center
\includegraphics[width=1.08 \textwidth]{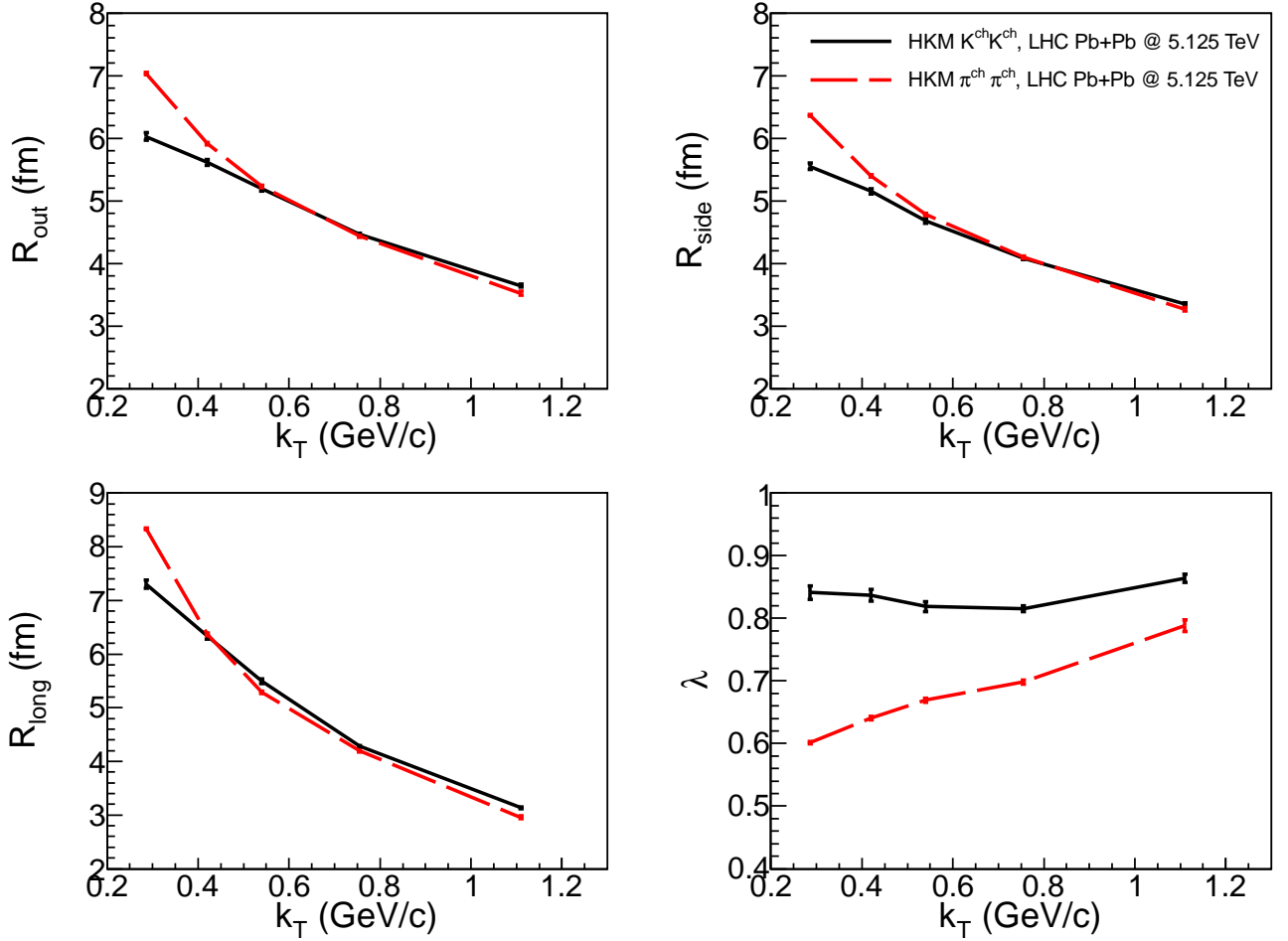}
\caption{Prediction for the $m_T$-dependence of $K^{ch}K^{ch}$ and $\pi^{-}\pi^{-}$ 3D interferometry radii for $\sqrt{s_{NN}}=5.125$~TeV Pb+Pb LHC collisions, $c=0-5\%$, $|\eta|<0.8$, $0.14<p_T<1.5$~GeV/$c$. For $\sqrt{s_{NN}}=2.76$~TeV Pb+Pb LHC case the corresponding radii values are $4-7\%$ lower for kaons and $2-4\%$ lower for pions.} 
\label{k-pi2}
\end{figure}

In Fig. \ref{k-pi2} we demonstrate the HKM predictions for full LHC energy $\sqrt{s_{NN}}=5.12$~TeV in $Pb+Pb$ collisions. The maximal initial energy density $\epsilon_0$ is chosen to reproduce the predicted mean charged particle multiplicity, taken from \cite{lhc513}. The initial flow is absent, $\alpha = 0$. The presented interferometry radii of kaons and pions demonstrate $k_T$-scaling for pion and kaon interferometry radii that starts from $k_T \approx 0.4$ GeV for {\it long}-radius and takes place for $k_T > 0.5$ GeV for ${\it out}$- and ${\it side}$- directions. The scaling is predicted also for Pb+Pb collisions at $\sqrt{s_{NN}}=2.76$~TeV, but corresponding radii values are $4-7\%$ lower for kaons and $2-4\%$ lower for pions than for full LHC energy.

\section{Conclusions}
The predictions of the hydrokinetic model are presented for the 1D and 3D interferometry radii of neutral and charged kaons in $Pb+Pb$ collisions at the LHC current energy $\sqrt{s_{NN}} = 2.76$ TeV and at the planning full energy $\sqrt{s_{NN}} = 5.125$. The magnitude of interferometry radii for charged and neutral kaons is very similar. The correlation functions do not have a purely Gaussian shape, especially for the 1D case resulting in the ``suppression'' parameter $\lambda$ being less than the intercept of the correlation functions and even in a decrease with transverse momentum of the pairs in the 1D case.  Partially because of this the parameter and radii practically are not  sensitive to effects of  $K^*$ and $\phi$ decays as was confirmed  by an analysis assuming the artificial exclusion of  the decays. A small contribution from these decays also sheds light on the near coincidence of the results for neutral and charged kaons.  Another feature of the predictions of the hydrokinetic model is absence of  $m_T$- scaling for pion and kaon interferometry radii. There is a significant violation of the scaling for the {\it long-} projection of the interferometry radii, that is caused, most likely, by strong transverse flow \cite{AkkSin}. However in the region of transverse pair momentum $k_T> 0.4-0.5$ GeV  the $k_T$-scaling is predicted by HKM. This is the result of the interplay of many different factors in the model, including the particular initial conditions. 

It will be very instructive to compare our predictions to data from ALICE experiment. The final aim is a quantitative characterization of the expansion and freeze-out dynamics of the fireball formed in central $Pb+Pb$ collisions at the LHC. 

\section{Acknowledgment}
Yu.S. is grateful to L.V. Malinina for fruitful discussions and to ExtreMe Matter Institute EMMI for support and visiting professor position.  
Iu.K. acknowledges the financial support by Hessian initiative for excellence (LOEWE) through the Helmholtz International Center for FAIR (HIC for FAIR).
The research was also carried out within the scope of the EUREA: European Ultra Relativistic Energies Agreement (European Research Group: ``Heavy ions at ultrarelativistic energies''), and is further supported by the National Academy of Sciences of Ukraine (Agreement 2014) and by the State fund for fundamental research of Ukraine (Agreement 2014).

\end{document}